\documentclass[12pt]{article}
\usepackage{epsfig}
\usepackage{amsfonts}
\usepackage{latexsym}
\usepackage{amsmath}
\usepackage{mathrsfs}
\usepackage{hyperref}
\usepackage{setspace}
\numberwithin{equation}{section}

\DeclareFontFamily{OT1}{rsfs}{}
\DeclareFontShape{OT1}{rsfs}{m}{n}{
<-7> rsfs5 <7-10> rsfs7 <10-> rsfs10}{}
\DeclareMathAlphabet{\mycal}{OT1}{rsfs}{m}{n}

\def\half{{1\over 2}}
\newcommand{\p}{\partial}

\newcommand{\bea}{\begin{eqnarray}}
\newcommand{\eea}{\end{eqnarray}}
\newcommand{\be}{\begin{equation}}
\newcommand{\ee}{\end{equation}}

  \makeatletter
  \let\over=\@@over \let\overwithdelims=\@@overwithdelims
  \let\atop=\@@atop \let\atopwithdelims=\@@atopwithdelims
  \let\above=\@@above \let\abovewithdelims=\@@abovewithdelims
  \makeatother

\begin{document}

\begin{titlepage}
\unitlength = 1mm
\vskip 1cm
\begin{center}

 { \LARGE {\textsc{Gravity Waves from Extreme-Mass-Ratio\\ \vskip.5cm  Plunges into Kerr Black Holes }}}

\vspace{1.8cm}
Shahar Hadar$^\S$, Achilleas P. Porfyriadis$^\dag$ and Andrew Strominger$^\dag$

\vspace{1cm}

{\it $^\S$ Racah Institute of Physics, Hebrew University, Jerusalem 91904, Israel}

\vspace{0.5cm}

{\it $^\dag$ Center for the Fundamental Laws of Nature, Harvard University,\\
Cambridge, MA 02138, USA}

\vskip 2cm

\begin{abstract}
Massive objects orbiting a near-extreme Kerr black hole quickly plunge into the horizon after passing the innermost stable circular orbit.  The plunge trajectory is shown to be related by a conformal map to a circular orbit. Conformal symmetry of the near-horizon region is then used to compute
the gravitational radiation produced during the plunge phase.
\end{abstract}

\vspace{1.0cm}

\end{center}
\end{titlepage}

\pagestyle{plain}
\setcounter{page}{1}
\newcounter{bean}
\baselineskip18pt


\setcounter{tocdepth}{2}

\tableofcontents

\section{Introduction}

General relativity implies that the high-redshift region very near the horizon of a near maximally-spinning Kerr black hole is governed by an infinite-dimensional conformal symmetry \cite{hep-th/9905099, 0809.4266}.  X-rays \cite{McClintock:2006xd} and iron lines \cite{Brenneman:2006hw} from such regions have already been observed, and the future may hold yet higher precision observations. It is of interest to explore any potential observational consequences of the  conformal symmetry. In a companion paper  \cite{Porfyriadis:2014fja}, the conformal symmetry was exploited  to compute gravity wave emission for an extreme-mass-ratio-inspiral within this near-horizon  region. This approximates the signal from a stellar mass object orbiting near an extreme supermassive Kerr black hole and is potentially observable at eLISA \cite{Finn:2000sy, Gair:2004iv, elisa}.  Once such an object passes the innermost stable circular orbit (ISCO), it plunges into the black hole. The plunge trajectory turns out to be related by a conformal map to the circular orbit. In this paper we use the conformal map to compute the gravitational radiation produced during this post-ISCO plunge into a near-extreme Kerr black hole. This complements the computation in \cite{Hadar:2009ip} of radiation produced during the post-ISCO plunge into a nonrotating Schwarzschild black hole.

In section 2 we set notation and briefly review the geometry of Kerr, near-horizon extreme Kerr (NHEK) and near-horizon near-extreme Kerr (near-NHEK).  Section 3 gives the conformal map from a circular orbit of a pointlike `star' in NHEK to a plunge trajectory in near-NHEK.  In section 4, as a warmup to the gravity case,  we couple a scalar field to the star. 4.1 computes the radiation production within near-NHEK using  a bulk gravity analysis. 4.2 computes the same process using the techniques of two-dimensional conformal field theory (CFT).  The results are shown to agree in subsection 4.3. In subsection 4.4 we turn to asymptotically flat near-extreme Kerr by reattaching the asymptotically flat region to near-NHEK. The resulting outgoing scalar radiation at flat future null infinity is computed. In 4.5 we derive the late-time quasinormal mode (QNM) decomposition. In section 5 all of these steps are repeated for the spin two case of gravitational radiation.

\section{Kerr, NHEK and near-NHEK}
The Kerr metric in Boyer-Lindquist coordinates reads ($G=\hbar=c=1$):
\be\label{Kerr metric}
ds^2=-\frac{\Delta}{\hat{\rho}^2}\left(d\hat{t}-a\sin^2\theta d\hat{\phi}\right)^2+\frac{\sin^2\theta}{\hat{\rho}^2}\left((\hat{r}^2+a^2)d\hat{\phi}-a d\hat{t}\right)^2+\frac{\hat{\rho}^2}{\Delta}d\hat{r}^2+{\hat{\rho}}^2 d\theta^2\,,
\ee
\be
\Delta=\hat{r}^2-2M\hat{r}+a^2, \qquad \hat{\rho}^2=\hat{r}^2+a^2\cos^2\theta\,.\nonumber
\ee
It is labeled by the mass $M$ and angular momentum $J=aM$. The horizons are located at:
\be
r_\pm=M\pm\sqrt{M^2-a^2}\,.
\ee
The Hawking temperature, angular velocity of the horizon, and Bekenstein-Hawking entropy are:
\be
T_H=\frac{r_+-M}{4\pi Mr_+}\,,\quad \Omega_H=\frac{a}{2Mr_+} \,,\quad S_{BH}=2\pi Mr_+\,.
\ee

Extreme Kerr is characterized by $a=M$ so that angular momentum takes the maximal value $J=M^2$. 
Near extremality one has
\be
\kappa\equiv\sqrt{1-\left(\frac{a}{M}\right)^2}\ll 1\,.
\ee
In this case there is a long throat to the horizon of the black hole and we can derive a regular throat geometry by zooming in to the horizon (Appendix \ref{app on throat geometries}). Defining
\be\label{near-NHEK coords}
r=\frac{\hat r-r_+}{r_+},\qquad t=\frac{\hat t}{2M},\qquad\phi=\hat\phi-\frac{\hat t}{2M}\,,
\ee
the near-NHEK geometry \cite{0907.3477} is obtained by scaling $r\sim\kappa$ to zero:
\be\label{nearNHEK}
ds^2=2M^2\Gamma(\theta)\left[-r(r+2\kappa)dt^2+\frac{dr^2}{r(r+2\kappa)}+d\theta^2+\Lambda(\theta)^2(d\phi+(r+\kappa)dt)^2\right]\,,
\ee
where
\be
\Gamma(\theta) = {1+\cos^2\theta\over 2} \ , \quad \Lambda(\theta) = {2\sin\theta\over 1 + \cos^2\theta}\,.
\ee
The NHEK geometry \cite{hep-th/9905099} is obtained by scaling $\kappa$ to zero first:
\be\label{NHEK}
ds^2=2M^2\Gamma(\theta)\left[-R^2dT^2+\frac{dR^2}{R^2}+d\theta^2+\Lambda(\theta)^2(d\Phi+R dT)^2\right]\,.
\ee

\section{Mapping from near-NHEK plunge to NHEK orbit}\label{Section: Transn}

\begin{figure}[!hb]
\includegraphics[angle=90, width=1.7in]{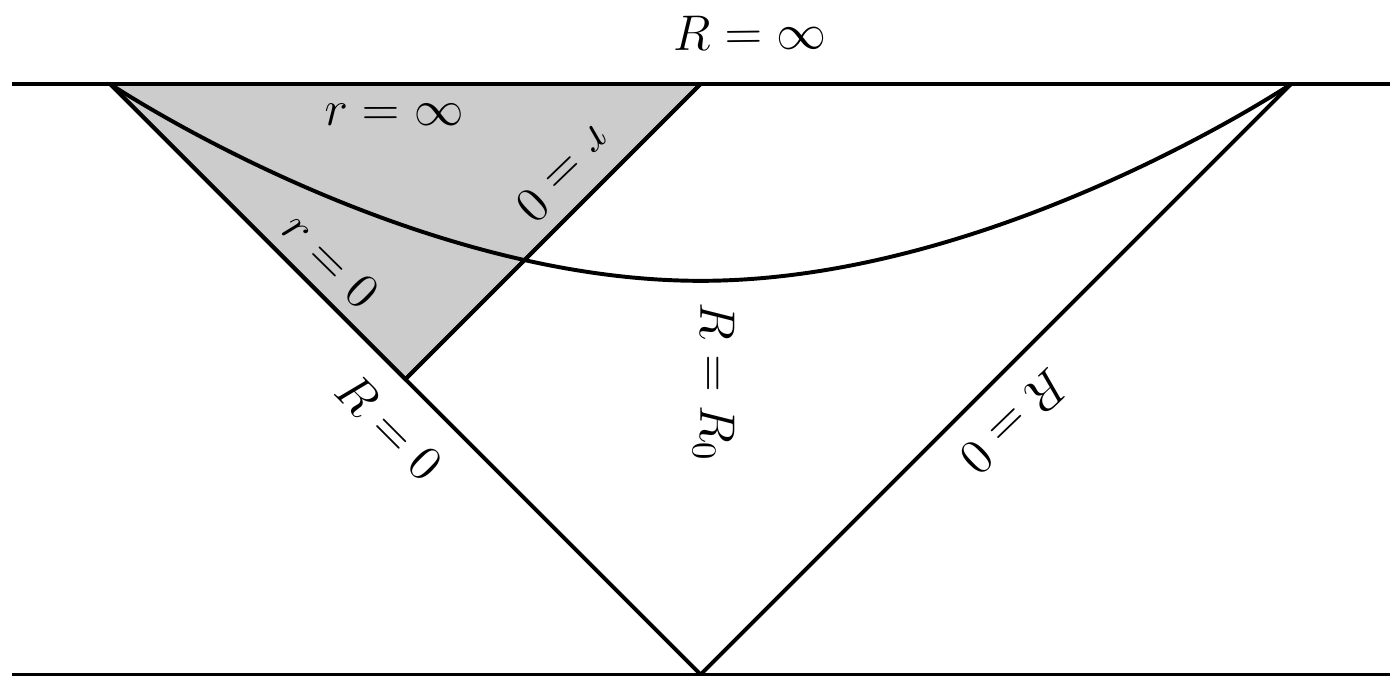}\centering
\caption{Penrose diagram of the throat geometry. The large wedge (bounded by $R=0$ and $R=\infty$) is NHEK. The small shaded wedge (bounded by $r=0$ and $r=\infty$) is near-NHEK. The line $R=R_0$ is the circular orbit in NHEK which in near-NHEK is seen to plunge into the future horizon at $r=0$.}\label{diagram}
\end{figure}

We wish to study radiation sourced by a particle that spirals off the ISCO in near-NHEK and plunges into the horizon. As shown in Appendix \ref{app on geodesics} this equatorial plunge orbit is given by:
\begin{eqnarray}
t(r)&=&\frac{1}{2\kappa}\ln\frac{1}{r(r+2\kappa)}+t_0\,,\label{t(r) plunge}\\
\phi(r)&=&\frac{3r}{4\kappa}+\frac{1}{2}\ln\frac{r}{r+2\kappa}+\phi_0\label{phi(r) plunge}\,,
\end{eqnarray}
and has energy and angular momentum (per unit rest mass):
\be
e=0\,,\qquad l=2M/\sqrt{3}\,.
\ee
This orbit is very special in that it is a Killing flow. The corresponding Killing vector becomes manifest when we make the following transformation to NHEK \cite{Maldacena:1998uz, Spradlin:1999bn, Amsel:2009ev, Dias:2009ex}:
\begin{eqnarray}\label{transn}
T&=&-e^{-\kappa t}\frac{r+\kappa}{\sqrt{r(r+2\kappa)}}\,,\nonumber\\
R&=&\frac{1}{\kappa}e^{\kappa t}\sqrt{r(r+2\kappa)}\,,\\
\Phi&=&\phi-\half\ln{r\over r+2\kappa}\,.\nonumber
\end{eqnarray}
This is illustrated in Figure \ref{diagram}. The transformation takes the near-NHEK equatorial plunge geodesics (\ref{t(r) plunge}, \ref{phi(r) plunge}) to the equatorial circular orbits in NHEK:
\begin{eqnarray}\label{plunge in NHEK}
R&=&R_0\,,\label{NHEK R=R0 plunge}\\
\Phi(T)&=&-\frac{3}{4}R_0\,T+\Phi_0\,, \label{NHEK Phi(T) plunge}
\end{eqnarray}
where,
\be
R_0=e^{\kappa t_0}/\kappa\,,\quad \Phi_0=\phi_0-3/4\,.
\ee
Note that (\ref{NHEK R=R0 plunge}-\ref{NHEK Phi(T) plunge}) is precisely what we would call an equatorial plunge trajectory in NHEK, that is, the geodesic with energy and angular momentum those of the marginally stable circular orbits of NHEK: $E=0, L=2M/\sqrt{3}$ (Appendix \ref{app on geodesics}).

The NHEK problem has a manifest Killing symmetry with respect to $\chi=\p_T-(3/4)R_0\p_\Phi$ and has been fully solved, both from the gravity and the CFT point of view, in \cite{Porfyriadis:2014fja}. In this paper we will thus solve the near-NHEK plunge problem by performing the bulk diffeomorphism \eqref{transn} on the gravity computation and the boundary conformal transformation,
\bea
T&=&-e^{-\kappa t}\,,\label{conf transn}\\
\Phi&=&\phi\,,
\eea
on the CFT computation of \cite{Porfyriadis:2014fja}. We will see that the results remain in perfect agreement.
Note that this conformal transformation is not among the $SL(2,R)$ isometries and hence this agreement tests the full infinite-dimensional conformal symmetry.

\section{Scalar radiation from a plunging star}
In this section we consider  a massless scalar field $\Psi$ coupled to a source $\mathcal{S}$ via the interaction:
\be
S_I=-4\pi\lambda\int d^4x\sqrt{-g}\,\Psi(x)\mathcal{S}(x)\,,
\ee
where $\lambda$ is a coupling constant. The source due to a pointlike `star' on a geodesic $x_*^\mu(\tau)$ is the covariant delta function integrated along the worldline of the star: \be\mathcal{S}(x)=-\int d\tau (-g)^{-1/2}\delta^{(4)}(x-x_*(\tau))\,.\ee
In this case the wave equation becomes
\be\label{wave eqn}
\square\, \Psi(x)=-4\pi\lambda\,\int d\tau (-g)^{-1/2}\delta^{(4)}(x-x_*(\tau))\,.
\ee

\subsection{Gravity analysis}\label{section: scalar gravity analysis}
In this subsection we solve the wave equation \eqref{wave eqn} for the near-NHEK plunge and compute the flux across the future horizon. The solution for a particle on the geodesic (\ref{NHEK R=R0 plunge}-\ref{NHEK Phi(T) plunge}) in NHEK with boundary conditions ingoing at the horizon and Neumann at the boundary is (setting $\Phi_0=0$) \cite{Porfyriadis:2014fja}:
\be
\Psi=\sum_{\ell,m}e^{im\left(\Phi+3R_0T/4\right)}S_\ell(\theta) R_{\ell m}(R)\,,
\ee
where $S_\ell$ are the spheroidal harmonics obeying
\be {1\over\sin\theta}\p_\theta(\sin\theta\,\p_\theta S_\ell)+\left(K_\ell-\frac{m^2}{\sin^2\theta}-\frac{m^2}{4}\sin^2\theta\right)S_\ell= 0\,,\label{angular eqn}
\ee
and $R_{\ell m}$ is given by:
\be\label{radial soln}
R_{\ell m}(R)={1\over W}\left[X\,\Theta(R_0-R) W_{im,h-\half}\left(-{2i\Omega}/{R}\right) +Z\,\Theta(R-R_0) M_{im,h-\half}\left(-{2i\Omega}/{R}\right)\right]\, ,
\ee
where $W_{k,\mu}$ and $M_{k,\mu}$ are Whittaker functions and
\begin{eqnarray*}
\Omega&=& -{3\over 4}mR_0\,,\cr
h&=&\half+\sqrt{1/4+K_\ell-2m^2}\,,\cr
X&=&-\frac{\sqrt{3}\lambda R_0}{2M} S_\ell(\pi/2)  M_{im,h-\half}\left({3im}/{2}\right)\,,\cr
Z&=&-\frac{\sqrt{3}\lambda R_0}{2M}S_\ell(\pi/2) W_{im,h-\half}\left({3im}/{2}\right)\,,\cr
W&=&2i\Omega\frac{\Gamma(2h)}{\Gamma(h-im)}\,.
\end{eqnarray*}
This solution has the asymptotic behaviors:
\begin{eqnarray}
\Psi(R\to 0)&=&\sum_{\ell,m} e^{im(\Phi+3R_0T/4)}\,S_\ell(\theta)\,\frac{ X}{W} (-2i\Omega)^{im}\, R^{-im}e^{-3imR_0/4R}\,,\\
\Psi(R\to \infty)&=&\sum_{\ell,m}e^{im(\Phi+3R_0T/4)}\,S_\ell(\theta)\,\frac{ Z}{W} (-2i\Omega)^{h} \,R^{-h}\,.
\end{eqnarray}

As explained in section \ref{Section: Transn}, the solution to the near-NHEK plunge problem, with the same boundary conditions, may be obtained by transforming the above NHEK solution using the transformation \eqref{transn}. That is to say, the $\Psi$ given above, transformed via \eqref{transn}, will be the solution to the wave equation \eqref{wave eqn} in near-NHEK with a source due to a plunging particle on the trajectory (\ref{t(r) plunge}-\ref{phi(r) plunge}).

Holding $t,\phi$ fixed, from the transformation \eqref{transn} we see that
\bea
R&\approx& \sqrt{{2\over \kappa}}\,e^{\kappa t}\,r^{1/2}\to 0 \qquad \textrm{for $r\to 0$}\,,\label{R(r->0)}\\
R&\approx& {e^{\kappa t}\over \kappa}\, r\to\infty \qquad\qquad\,\,\, \textrm{for $r\to \infty$}\,,\label{R(r->infty)}
\eea
and
\begin{eqnarray}
e^{im(\Phi+3R_0T/4)}\,(r\to 0)&=& (2\kappa)^{im/2} e^{im\phi} r^{-im/2} \exp\left[-\frac{3imR_0}{4}\sqrt{\frac{\kappa}{2r}}\,e^{-\kappa t}\right] \label{e^im(Phi+...)(r->0)} \,,\\
e^{im(\Phi+3R_0T/4)}\,(r\to \infty)&=& e^{im\phi}\exp\left[-\frac{3imR_0}{4}\,e^{-\kappa t}\right] \,. \label{e^im(Phi+...)(r->infty)}
\end{eqnarray}
We thus find:
\bea
\Psi(r\to 0)&=&\sum_{\ell,m} \frac{ X}{W} \kappa^{im}(-2i\Omega)^{im}\,e^{im\phi}\,S_\ell(\theta)\, r^{-im}e^{-im\kappa t}\exp\left[-\frac{3imR_0}{2}\sqrt{\frac{\kappa}{2r}}\,e^{-\kappa t}\right]\,,\\
\Psi(r\to \infty)&=&\sum_{\ell,m} \frac{ Z}{W} \kappa^h (-2i\Omega)^{h}\,e^{im\phi}\,S_\ell(\theta)\, r^{-h}e^{-h\kappa t}\exp\left[-\frac{3imR_0}{4}\,e^{-\kappa t}\right]\,.
\eea
In terms of the Fourier decomposition\footnote{The relevant Fourier transforms are:
\[
F\left(e^{-im\kappa t}\exp\left[-\frac{3imR_0}{2}\sqrt{\frac{\kappa}{2r}}\,e^{-\kappa t}\right]\right)= {1\over\sqrt{2\pi}\kappa}\left(\frac{2}{\kappa}\right)^{i{m\over 2}-i{\omega\over2\kappa}} \left(\frac{3imR_0}{2}\right)^{-im+i{\omega\over\kappa}}\Gamma\left(im-i{\omega\over\kappa}\right) r^{i{m\over 2}-i{\omega\over 2\kappa}}\,.
\]
\[
F\left(e^{-h\kappa t}\exp\left[-\frac{3imR_0}{4}\,e^{-\kappa t}\right]\right)= {1\over\sqrt{2\pi}\kappa}\left(\frac{3imR_0}{4}\right)^{-h+i{\omega\over\kappa}}
\Gamma\left(h-i{\omega\over\kappa}\right)\,.
\]
}:
\begin{eqnarray}
\Psi(r\to 0)&=&{1\over\sqrt{2\pi}}\int d\omega\sum_{\ell,m} I\,e^{-i\omega t}\,e^{im\phi}\,S_\ell(\theta) r^{-{i\over 2}(m+{\omega/ \kappa})}\,,\label{Psi(r->0 soln)}\\
\Psi(r\to \infty)&=&{1\over\sqrt{2\pi}}\int d\omega\sum_{\ell,m}N\,e^{-i\omega t}\,e^{im\phi}\,S_\ell(\theta)r^{-h}\,,\label{Psi(r->infty soln)}
\end{eqnarray}
where
\bea
I&=&{1\over\sqrt{2\pi}}2^{{i\over 2}(m-{\omega/\kappa})}\frac{ X}{W} \kappa^{-1+{i\over 2}(m+{\omega/ \kappa})} (-2i\Omega)^{i{\omega/\kappa}} \Gamma(im-i{\omega/\kappa})\,,\\
N&=&{1\over\sqrt{2\pi}}2^{h-i{\omega/ \kappa}}\frac{ Z}{W} \kappa^{h-1} (-2i\Omega)^{i{\omega/\kappa}} \Gamma(h-i{\omega/\kappa})\,.\label{N expression}
\eea
The Klein-Gordon particle number flux is defined as:
\be
\mathcal{F}=\int\sqrt{-g}J^r\,d\theta d\phi\,,\quad J^\mu={i\over 8\pi} \left(\Psi^*\nabla^\mu\Psi-\Psi\nabla^\mu\Psi^*\right)\,.
\ee
The near-NHEK solution (\ref{Psi(r->0 soln)}-\ref{Psi(r->infty soln)}) obeys ingoing boundary conditions at the horizon and Neumann at the boundary. Thus for real $h$ the Klein-Gordon particle number flux vanishes at the boundary. Finally, the horizon flux, for $\omega>0\,, m>0$, is:
\bea
\mathcal{F}_{\ell m \omega}&=&{M^2\over 4\pi\kappa}\left|\frac{ X}{W} \right|^2 \frac{m+\omega/\kappa}{m-\omega/\kappa} \frac{e^{-\pi\omega/\kappa}}{\sinh\pi(m-\omega/\kappa)}\\
&\approx&{M^2\over 2\pi\kappa} \left|\frac{ X}{W} \right|^2 e^{-2\pi\omega/\kappa} e^{\pi m}\,, \label{ff}
\eea
where in the last line we have used $\kappa\ll 1$.

\subsection{CFT analysis}
In this subsection we derive the flux formula (\ref{ff}) from the CFT representation of gravity in near-NHEK as a 2D CFT at temperatures $T_L={1 / 2\pi},~~T_R={ \kappa / 2\pi}$ and an angular potential \cite{0907.3477}. As shown in \cite{Porfyriadis:2014fja}, the addition of  a star to NHEK is dual to a certain deformation of the CFT.  The latter is described by the deformation of the two-dimensional action $S_{CFT}$:
\be\label{CFT action deformation}
S = S_{CFT}+\sum_{\ell}\int d\Phi\, dT\, J_{\ell}(\Phi,T)\mathcal{O}_{\ell}(\Phi,T)\,,
\ee
where $\mathcal{O}_\ell$ are CFT operators with  left and right weights $h$.\footnote{To avoid index clutter we suppress here and in the following the $\ell$ subscript on $h$.} For a star on the orbit (\ref{NHEK R=R0 plunge}-\ref{NHEK Phi(T) plunge}) it was shown that \cite{Porfyriadis:2014fja}
\be\label{source from paper1}
J_{\ell}(\Phi,T)=\sum_m \frac{X}{W}C\, e^{im(\Phi+3R_0T/4)}\,,
\ee
where  $C=(-2i\Omega)^{1-h}{\Gamma(2h-1)}/{\Gamma(h-im)}$.
Since $\mathcal{O}_\ell$ carry left and right weights $h$ it follows by conformal invariance of \eqref{CFT action deformation} that $J_\ell$ carry left and right weights $1-h$. As a result, the conformal transformation \eqref{conf transn} gives:
\be
J_\ell(\phi,t)=\kappa^{1-h} e^{(h-1)\kappa t}\sum_m \frac{X}{W}C\, e^{im(\phi-3R_0e^{-\kappa t}/4)}\,.
\ee
The Fourier decomposition is:
\be\label{J_ell(phi,t)}
J_\ell(\phi,t)={1\over\sqrt{2\pi}}\int d\omega\sum_{\ell,m} J_{\ell m \omega}\,e^{-i\omega t}\,e^{im\phi}\,,
\ee
where
\be
J_{\ell m \omega}={1\over\sqrt{2\pi}}\,\frac{X}{W}C\, \kappa^{-h} \left({3imR_0\over 4}\right)^{h-1+i\omega/\kappa}\Gamma(1-h-i\omega/\kappa)\,.
\ee
The deformed action \eqref{CFT action deformation} may then be written:
\be
S=S_{CFT}+{1\over\sqrt{2\pi}}\int d\omega\sum_{\ell,m}\int d\phi\, dt\, J_{\ell m \omega}e^{im\phi-i\omega t}\mathcal{O}_\ell(\phi,t)\,.
\ee

The time dependent sources pump both left and right energy into the CFT.
The leading-order transition rate out of the thermal state is given by Fermi's golden rule \cite{hep-th/9702015, hep-th/9706100, Porfyriadis:2014fja}:
\be\label{total rate}
\mathcal{R}=\int d\omega \sum_{\ell, m}|J_{\ell m \omega}|^2 \int d\phi\,dt\,  e^{-im\phi+ i\omega t}G(\phi,t)\,,
\ee
where $G(\phi,t)=\langle \mathcal{O}^\dagger(\phi,t)\mathcal{O}(0,0)\rangle_{T_L,T_R}$ is the finite temperature two point function of the  CFT.  Using $T_L={1 \over 2\pi},~~T_R={ \kappa\over 2\pi}$, for $\omega>0\,,m>0$  with the appropriate $i\epsilon$ prescription, the Fourier transform of the two point function is:
\bea\label{partial wave rate}
\mathcal{R}_{\ell m \omega}&=&C_\mathcal{O}^2\, \frac{1}{\Gamma(2h)^2}\, |J_{\ell m \omega}|^2\,\kappa^{2h-1} e^{\pi\omega/\kappa}e^{\pi m}|\Gamma(h+i\omega/\kappa)|^2 |\Gamma(h+im)|^2\\
&\approx& C_\mathcal{O}^2\, \frac{8\pi}{2^{2h}(2h-1)^2 \kappa} \left|\frac{ X}{W} \right|^2 e^{-2\pi\omega/\kappa} e^{\pi m}\,,
\eea
where $C_\mathcal{O}^2$ is the operator normalization and in the last line we have used $\kappa\ll 1$.

\subsection{Gravity/CFT matching}
With the normalization found in \cite{Porfyriadis:2014fja},
\be \label{dd} C_{\mathcal O}= {2^{h-1}(2h-1)\over  2\pi}M\,,\ee
we have the exact match of the gravity and CFT computations:
\be
\mathcal{R}_{\ell m \omega}={M^2\over 2\pi\kappa} \left|\frac{ X}{W} \right|^2 e^{-2\pi\omega/\kappa} e^{\pi m}={\cal F}_{\ell m \omega}\,.
\ee

\subsection{Reattaching the asymptotically flat region}

In the previous subsections we considered scalar radiation production in near-NHEK. Boundary conditions were imposed such that all radiation was (for real $h$) reflected off of the asymptotic throat boundary and ultimately falls into the future horizon. Here we reattach the asymptotically flat region and compute the outgoing radiation at future null infinity.

Consider the scalar field on Kerr expanded in modes:
\be\label{Psi in Kerr expansion}
\Psi={1\over\sqrt{2\pi}}\int d\hat\omega\sum_{\ell, m} e^{-i\hat\omega \hat t}e^{im\hat\phi}\hat S_\ell(\theta) \hat R_{\ell m \hat\omega}(\hat r)\,.
\ee
The wave equation separates into the spheroidal angular equation,
\be
{1\over\sin\theta}\p_\theta(\sin\theta\,\p_\theta \hat S_\ell)+\left(\hat K_\ell-\frac{m^2}{\sin^2\theta}-a^2\hat\omega^2\sin^2\theta\right)\hat S_\ell=0\,,
\ee
and a radial equation which in the coordinates \eqref{nearNHEK} reads:
\be\label{radial x eqn}
r(r+\tau_H)\hat R_{\ell m \hat\omega}'' + (2r + \tau_H)\hat R_{\ell m \hat\omega}' + V \hat R_{\ell m \hat\omega} = \hat T_{\ell m \hat\omega}\,,
\ee
with
\be
V = \frac{(r_+\hat\omega r^2 + 2 r_+ \hat\omega r + n\tau_H/2)^2}{r(r+\tau_H)} + 2am\hat\omega - \hat K_\ell\,.
\ee
Here $\hat K_\ell$ is the separation constant and,
\be
\tau_H\equiv{r_+-r_-\over r_+}\,,\qquad n\equiv4M\frac{\hat\omega-m\Omega_H}{\tau_H}\,.
\ee

In the near extremal, near superradiant bound regime,
\be\label{regime}
\tau_H\ll 1\quad\textrm{and}\quad n\tau_H\ll 1\,,
\ee
we have to leading order:
\be\label{relations1}
\tau_H=2\kappa \ , \ 2r_+\hat\omega=2a\hat\omega=m \ , \ \hat K_\ell=K_\ell \ , \ \hat S_\ell=S_\ell \ , \ 2M\hat\omega-m=(n-m)\kappa\,.
\ee
Then, identifying
\be\label{relations2}
\omega=(n-m)\kappa\,,
\ee
we can match a near-NHEK solution containing the source with a far Kerr vacuum solution as follows. For $r\gg max(\tau_H, n\tau_H)$ equation \eqref{radial x eqn} becomes the vacuum far equation in extreme Kerr and the solution that is purely outgoing at null infinity behaves as \cite{Porfyriadis:2014fja}:
\begin{eqnarray}
\hat R_{\ell m \hat\omega}^{far}(r\to \infty)&=&Q\,\frac{\Gamma(2-2h)}{\Gamma(1-h+im)}(im)^{h-1+im} \times \label{R^far at infinity}\\
 &&~~~\times\left[1- \frac{(-im)^{2h-1}}{(im)^{2h-1}}\frac{\sin\pi(h+im)}{\sin\pi(h-im)}\right]r^{-1+im}e^{imr/2} \,,\notag\\
\hat R_{\ell m \hat\omega}^{far}(r\to 0)&=&P\, r^{h-1}+Q\,r^{-h}\,,\label{R^far at throat}
\end{eqnarray}
where
\be\label{P/Q}
{P\over Q}=-(-im)^{2h-1}\frac{\Gamma(2-2h)}{\Gamma(2h)}\frac{\Gamma(h-im)}{\Gamma(1-h-im)}\,.
\ee
For $r\ll 1$ equation \eqref{radial x eqn} becomes the near-NHEK equation,
\be\label{nearNHEK radial eqn}
r(r+2\kappa)\hat R_{\ell m \hat\omega}'' + 2(r + \kappa)\hat R_{\ell m \hat\omega}' + \left[\frac{(mr+n\kappa)^2}{r(r+2\kappa)}+m^2-K_\ell\right]\hat R_{\ell m \hat\omega} = \hat T_{\ell m \hat\omega}\,,
\ee
with the source $\hat T_{\ell m \hat\omega}$ due to the plunging star on (\ref{t(r) plunge}-\ref{phi(r) plunge}). In section \ref{section: scalar gravity analysis} we found the solution to this equation with ingoing boundary conditions at the horizon and Neumann at the boundary of the throat (equations (\ref{Psi(r->0 soln)}-\ref{Psi(r->infty soln)})). The latter implies no flux leaking outside the throat. However, here we wish to allow the necessary flux leak out in such a way that we can match the large $r$ behavior of the solution to \eqref{nearNHEK radial eqn} with the small $r$ behavior of $\hat R_{\ell m \hat\omega}^{far}$ in \eqref{R^far at throat}. This is done by adding to the solution of section \ref{section: scalar gravity analysis} an ingoing at the horizon homogeneous solution of \eqref{nearNHEK radial eqn},
\be
\hat R_{in}^{near}=r^{-in/2}\left({r\over 2\kappa}+1\right)^{i(n/2-m)}{_2}{F}{_1}\left(h-im\,,1-h-im\,,1-in\,,-{r\over 2\kappa}\right)\,,
\ee
with the appropriate amplitude to match the Dirichlet mode of \eqref{R^far at throat}. Doing so fixes the amplitude $Q$ according to:
\be
Q=2MN\left[1-(-2im\kappa)^{2h-1}\frac{\Gamma(1-2h)^2}{\Gamma(2h-1)^2}\frac{\Gamma(h-im)^2}{\Gamma(1-h-im)^2} \frac{\Gamma(h-i(n-m))}{\Gamma(1-h-i(n-m))}\right]^{-1}\,,
\ee
where $N$ is given in \eqref{N expression}. Plugging into equation \eqref{R^far at infinity} we obtain the waveform at future null infinity as a function of $\ell,m$ and $\hat \omega$:
\begin{eqnarray} \label{tf}
&&\hat R_{\ell m \hat\omega}^{far}(r\to \infty) =\frac{4\lambda}{\sqrt{6\pi}}(-1)^{-h}(2\kappa)^{h-1}(3R_0/4)^{i(n-m)}\times \label{R^far at infinity final}\\
&&~~~~\times \frac{ S_\ell(\pi/2) W_{im,h-\half}\left({3im}/{2}\right) (im)^{h-2+in} e^{\pi m} {(1-2h)\Gamma(h-im)^2}/{\Gamma(2h)^2} }{ \frac{1}{\Gamma(h-i(n-m))} -(-2im\kappa)^{2h-1}\frac{\Gamma(1-2h)^2}{\Gamma(2h-1)^2}\frac{\Gamma(h-im)^2}{\Gamma(1-h-im)^2} \frac{1}{\Gamma(1-h-i(n-m))} }\, r^{-1+im}e^{imr/2}\,, \notag
\end{eqnarray}
for $m>0$.

\subsection{Quasinormal mode decomposition}
The waveform at future infinity as a function of time rather than frequency is obtained by plugging \eqref{tf} into \eqref{Psi in Kerr expansion} and integrating over frequencies. It is well known that for appropriately late times the waveform is dominated by the quasinormal mode contribution, which is determined by the residues of $\hat R_{\ell m \hat\omega}^{far}(r\to \infty)$ at its poles. From \eqref{R^far at infinity final} the poles are located where the denominator,
\be
\mathcal{D}\equiv \frac{1}{\Gamma(h-i(n-m))}- (-2im\kappa)^{2h-1}\frac{\Gamma(1-2h)^2}{\Gamma(2h-1)^2}\frac{\Gamma(h-im)^2}{\Gamma(1-h-im)^2} \frac{1}{\Gamma(1-h-i(n-m))}\,,
\ee
vanishes. These QNMs of near extreme Kerr were first studied for the case $n\to\infty$ in \cite{Detweiler:1980gk} and more recently for $n=\textrm{finite}$ in \cite{Hod:2008zz}. Here we follow the latter because this case gives the most long lived resonances. In order to have $\mathcal{D}=0$, the two different terms need to be of the same magnitude as $\kappa\to 0$ which happens when $\Gamma(h-i(n-m))$ itself is close to a pole. Accordingly, we write the following ansatz for the QNM frequency:
\be
n=m-i(N+h-\epsilon \eta)\,, \label{nQNM}
\ee
where $\epsilon\equiv(-2im\kappa)^{2h-1} \ll 1$ for real $h$, and $N=0,1,2,\ldots$ is the overtone number which enumerates the different poles. Then we can approximate,
\be\label{gamma function approximation}
\Gamma(h-i(n-m)) = \Gamma(-N+\epsilon \eta) \approx \frac{(-1)^N}{N!\, \epsilon\, \eta}\,,
\ee
so that $\mathcal{D}$ near the QNM frequencies may be written as:
\be
\mathcal{D} \approx \epsilon \, (-1)^N \, N! \, \left( \eta - \eta_{QNM} \right)\,, \label{D near qnms}
\ee
with
\be
\eta_{QNM} =  \frac{\Gamma(1-2h)^2}{\Gamma(2h-1)^2}\frac{\Gamma(h-im)^2}{\Gamma(1-h-im)^2}\frac{(-1)^{N}}{N! \, \Gamma(1-2h-N)}\,. \label{etaQNM}
\ee
So all the poles in \eqref{R^far at infinity final} are simple poles in the lower complex $\hat\omega$ plane located at:
\be
\hat{\omega}_{N \ell m} = \frac{1}{2M} \left[ m - i \kappa (N + h) \right]\,. \label{qnm frequencies}
\ee
Then, by the residue theorem, the QNM contribution to the waveform at asymptotic flat infinity is given by:
\be
\Psi^{far}(r\to\infty) = \sum_{N,\, \ell,\, m} c_{N \ell m}\, e^{-i \hat{\omega}_{N \ell m} \hat{t}}\,  e^{i m \hat{\phi}}\, S_{\ell}(\theta)\,  r^{-1+im}  e^{imr/2}\,,\label{QNM waveform}
\ee
where the amplitudes are
\bea
c_{N \ell m} &=& \frac{\lambda}{\sqrt{3}M} \frac{(-1)^{N-h}}{N!} (2\kappa)^{h} (3R_0/4)^{N+h} S_\ell(\pi/2) W_{im,h-\half}\left({3im}/{2}\right)\times \notag \\
&&~~~\times (im)^{N+2h-2+im} e^{\pi m} {(1-2h) \Gamma(h-im)^2}/{\Gamma(2h)^2} \,.\label{qnm amplitudes}
\eea
Finally, performing the sum over $N$  we find:
\begin{eqnarray}
&&\Psi^{far}(r\to\infty) \notag\\
&&~=\sum_{\ell, m} \frac{3^{h-\half}\lambda}{2^h M} (-1)^{-h} e^{3im/4} S_\ell(\pi/2) W_{im,h-\half}\left({3im}/{2}\right) (im)^{2h-2+im} e^{\pi m} \frac{(1-2h) \Gamma(h-im)^2}{\Gamma(2h)^2} \times  \notag\\
&&\qquad\qquad\times   \exp \left[ -i\frac{m - i \kappa h}{2M}(\hat{t}-\hat{t}_0)-\frac{3im}{4 \kappa} e^{-\frac{\kappa}{2M} (\hat{t}-\hat{t}_0)} \right] e^{i m (\hat{\phi} - \hat{\phi}_0)}  S_{\ell}(\theta) \, r^{-1+im} e^{imr/2} \,, \label{QNM waveform after sum}
\end{eqnarray}
where we have used that $R_0=\kappa^{-1}e^{\kappa \hat t_0/(2M)}\,, \hat\phi_0={\hat t_0/ (2M)}+{3/ 4}$. We note that this QNM waveform is valid starting at times,
\be
\exp\left[-{\kappa\over 2M}(\hat{t}-\hat{t}_0)\right]\sim\kappa\,,
\ee
so that in \eqref{QNM waveform after sum} we have $\Psi^{far}\sim\kappa^h$.

\section{Gravitational radiation from a plunging star}
In this section we repeat the analysis for the realistic case of plunging stars coupled to gravity rather than scalars. It is conceptually similar but slightly more intricate computationally.

\subsection{Gravity analysis}\label{section: s=-2 gravity analysis}
In \cite{Porfyriadis:2014fja} the solution to the Teukolsky equation describing gravitational wave generation in NHEK due to a particle on (\ref{NHEK R=R0 plunge}-\ref{NHEK Phi(T) plunge}) was found to be:
\be
\psi^{(-2)}=\sum_{\ell,m}e^{im\left(\Phi+3R_0T/4\right)}S_\ell(\theta) R_{\ell m}(R)\,,
\ee
where $\psi^{(-2)}$ is a Newman-Penrose component of the Weyl tensor, $S_\ell$ are the spin-weighted spheroidal harmonics obeying
\be
{1\over\sin\theta}\p_\theta(\sin\theta\,\p_\theta S_\ell)+\left(K_\ell-\frac{m^2+s^2+2ms\cos\theta}{\sin^2\theta}- \frac{m^2}{4}\sin^2\theta-ms\cos\theta\right)S_\ell=0\,,\label{angular eqn s=-2}
\ee
and $R_{\ell m}$ is:
\be\label{radial soln s=-2}
R_{\ell m}(R)={1\over R_0^{-2s} W} \left[\mathcal{X}\,\Theta(R_0-R) \mathcal{W}(R)+ \mathcal{Z}\, \Theta(R-R_0) \mathcal{M}(R)\right]+a_2\delta(R-R_0)\,,
\ee
with
\bea
\mathcal{M}(R)&=& R^{-s}\,M_{im+s, h-\half}\left(-{2i\Omega}/{R}\right)\,,\notag\cr
\mathcal{W}(R)&=& R^{-s}\,W_{im+s, h-\half}\left(-{2i\Omega}/{R}\right)\,,\notag\cr
W&=& 2i\Omega\frac{\Gamma(2h)}{\Gamma(h-im-s)}\,,\notag\cr
\mathcal{X}&=& R_0\mathcal{M}'(R_0)(2sa_2-a_1-2a_2)+\mathcal{M}(R_0)(a_0-2sa_1-2sa_2+4s^2a_2-a_2V_s(R_0))\,,\cr
\mathcal{Z}&=&\mathcal{X}(\mathcal{M}\to\mathcal{W})\,,\notag\cr
V_s(R)&=& 2m^2-K_\ell+s(s+1)+\frac{2\Omega(m-is)}{R}+\frac{\Omega^2}{R^2}\,,\cr
a_0&=&\frac{m_0R_0^3}{16\sqrt{3}M^5}\left(40S+3imS-{9\over 8}m^2S+6mS'-16iS'-8S''\right)\,,\notag\\
a_1&=&\frac{m_0R_0^3}{16\sqrt{3}M^5}\left(8iS'-3imS-16S\right)\,,\notag\\
a_2&=&\frac{m_0R_0^3}{16\sqrt{3}M^5}2S\,.\notag
\end{eqnarray}
Here $S, S', S''$ denote $S_\ell(\pi/2), S'_\ell(\pi/2), S''_\ell(\pi/2)$ respectively and $s=-2$ is understood. This solution obeys ingoing boundary conditions at the horizon and Neumann at the boundary:
\begin{eqnarray}
\psi^{(-2)}(R\to 0)&=&\sum_{\ell,m} e^{im(\Phi+3R_0T/4)}\,S_\ell(\theta)\,\frac{\mathcal{X}}{R_0^{4}W} (-2i\Omega)^{im-2}\, R^{-im+4}e^{-3imR_0/4R}\,,\\
\psi^{(-2)}(R\to \infty)&=&\sum_{\ell,m}e^{im(\Phi+3R_0T/4)}\,S_\ell(\theta)\,\frac{\mathcal{Z}}{R_0^{4}W} (-2i\Omega)^{h} \,R^{-h+2}\,.
\end{eqnarray}

As in the scalar case, the solution to the Teukolsky equation in near-NHEK with a source due to a plunging particle on (\ref{t(r) plunge}-\ref{phi(r) plunge}) and with the same boundary conditions, may be obtained by transforming the above NHEK solution via equations \eqref{transn}. However, a minor additional step is in order here. In NHEK the solution above describes metric perturbations according to \cite{Porfyriadis:2014fja} $\psi^{(-2)}=(1-i\cos\theta)^4\delta\psi_4$ with the Weyl scalar $\psi_4\equiv C_{\mu\nu\rho\sigma}n^\mu \bar{m}^\nu n^\rho \bar{m}^\sigma$ defined with respect to the NHEK Kinnersley tetrad
\begin{eqnarray}
L^\mu &=&\left(\frac{1}{R^2},1,0,-{1\over R}\right)\,,\notag\\
N^\mu &=&\frac{1}{2M^2(1+\cos^2\theta)}\left(1,-R^2,0,-R\right)\,,\label{NHEK NP tetrad} \\
M^\mu &=&\frac{1}{\sqrt{2}M(1+i\cos\theta)}\left(0,0,1,\frac{i(1+\cos^2\theta)}{2\sin\theta}\right)\,.\notag
\end{eqnarray}
The transformation \eqref{transn} takes this NHEK Kinnersley tetrad to a near-NHEK tetrad in which the spin coefficient $\epsilon\neq 0$ but in which $l$ and $n$ are still aligned along the repeated principal null directions of the Weyl tensor. Thus to get to the near-NHEK Kinnersley tetrad,
\begin{eqnarray}
l^\mu &=&\left(\frac{1}{r(r+2\kappa)},1,0,-{r+\kappa\over r(r+2\kappa)}\right)\,,\notag\\
n^\mu &=&\frac{1}{2M^2(1+\cos^2\theta)}\left(1,-r(r+2\kappa),0,-(r+\kappa)\right)\,,\label{nearNHEK NP tetrad} \\
m^\mu &=&\frac{1}{\sqrt{2}M(1+i\cos\theta)}\left(0,0,1,\frac{i(1+\cos^2\theta)}{2\sin\theta}\right)\,,\notag
\end{eqnarray}
the transformation \eqref{transn} needs to be followed by a (Class III) rotation $l\to (R/r) l\,, n\to (r/R) n$ which rotates,
\be\label{psi null rotation}
\psi^{(-2)}\to \kappa^2 e^{-2\kappa t}\frac{r}{r+2\kappa}\,\psi^{(-2)}\,.
\ee
We do this primarily for later convenience when matching to the standard Kerr Teukolsky equation separated in the Kinnersley tetrad.

Using equations (\ref{R(r->0)}-\ref{e^im(Phi+...)(r->infty)}) followed by a Fourier transform
we get:
\begin{eqnarray}
\psi^{(-2)}(r\to 0)&=&{1\over\sqrt{2\pi}}\int d\omega\sum_{\ell,m} \mathcal{I}\,e^{-i\omega t}\,e^{im\phi}\,S_\ell(\theta) r^{-{i\over 2}(m+{\omega/\kappa})+2}\,,\label{psi-2(r->0) soln}\\
\psi^{(-2)}(r\to \infty)&=&{1\over\sqrt{2\pi}}\int d\omega\sum_{\ell,m}\mathcal{N}\,e^{-i\omega t}\,e^{im\phi}\,S_\ell(\theta)r^{-h+2}\,,\label{psi-2(r->infty) soln}
\end{eqnarray}
where
\bea
\mathcal{I}&=&{1\over\sqrt{2\pi}}2^{{i\over 2}(m-{\omega/\kappa})}\frac{\mathcal{X}}{R_0^{4}W} \kappa^{-1+{i\over 2}(m+{\omega/\kappa})} (-2i\Omega)^{i{\omega/\kappa}} \Gamma(-2+im-i{\omega/\kappa})\,,\\
\mathcal{N}&=&{1\over\sqrt{2\pi}}2^{h-i{\omega/ \kappa}}\frac{\mathcal{Z}}{R_0^{4}W} \kappa^{h-1} (-2i\Omega)^{i{\omega/\kappa}} \Gamma(h-i{\omega/\kappa})\,.\label{N expression s}
\eea
For $\omega>0, m>0$, the graviton number flux at the horizon, to leading order in $\kappa$, is \cite{Teukolsky:1974yv}:
\bea
\mathcal{F}_{\ell m \omega}&=&\frac{8 M^{10}}{\pi\kappa}\left|\frac{\mathcal{X}}{R_0^{4}W}\right|^2 \frac{e^{-2\pi\omega/\kappa}}{|\mathcal{C}|^2} e^{\pi m}\,,\label{F_ell m omega gravitons}\\
|{\mathcal C}|^2&\equiv &\left((K_\ell-m^2)^2+m^2\right)\left((K_\ell-m^2-2)^2+9m^2\right)\,.\nonumber
\eea

\subsection{Gravity/CFT matching}
The source corresponding to the NHEK problem, as obtained from the leading term of the appropriate Hertz potential at the boundary, is given by \cite{Porfyriadis:2014fja}:
\be
J_\ell(\Phi,T)=\sum_m {4\over \mathcal{C}}{\mathcal{X}\over R_0^{-2s} W}\frac{(-2i\Omega)^{1-h}\Gamma(2h-1)}{\Gamma(h-im-s)} \, e^{im(\Phi+3R_0T/4)}\,.
\ee
Now $\mathcal{O}_\ell$ carry weights \cite{Hartman:2009nz} $h_R=h\,, h_L=h-s$ so that conformal invariance implies $J_\ell$ carry right weight $1-h$ and left weight $1-h+s$. Thus as in the scalar case, after the conformal transformation we find $J_\ell(\phi, t)$ given as in \eqref{J_ell(phi,t)} with:
\be
J_{\ell m \omega}={2^{3-h-i\omega/\kappa} \over \sqrt{2\pi}\, \mathcal{C}} \kappa^{-h} {\mathcal{X}\over R_0^{-2s} W}(-2i\Omega)^{i\omega/\kappa}\Gamma(1-h-i\omega/\kappa)\frac{\Gamma(2h-1)}{\Gamma(h-im-s)}\,.
\ee
Using the above together with the operator normalization from \cite{Porfyriadis:2014fja},
\be
{C_{\cal O}}=\frac{2^{h-1}M^5}{2\pi}\frac{\sqrt{\Gamma(2h+4)\Gamma(2h)}}{\Gamma(2h-1)}\,,
\ee
and plugging into Fermi's golden rule reproduces, to leading order in $\kappa$, the rate \eqref{F_ell m omega gravitons}.

\subsection{Reattaching the asymptotically flat region}
Consider the full Teukolsky equation in Kerr for $\hat\psi^{(-2)}\equiv (\hat r-ia\cos\theta)^4\delta\hat\psi_4$ expanded in modes:
\be
\hat\psi^{(-2)}={1\over\sqrt{2\pi}}\int d\hat\omega\sum_{\ell, m} e^{-i\hat\omega \hat t}e^{im\hat\phi}\hat S_\ell(\theta) \hat R_{\ell m \hat\omega}(\hat r)\,.
\ee
In the near extremal, near superradiant regime \eqref{regime}, given the relations (\ref{relations1}-\ref{relations2}), we can match, across the entrance to the throat, an ingoing at the horizon solution of the near-NHEK equation containing the source with a far Kerr vacuum solution that is purely outgoing at null infinity. The desired far solution is given by \cite{Porfyriadis:2014fja}:
\begin{eqnarray}
\hat R_{\ell m \hat\omega}^{far}(r\to \infty) &=& Q\,\frac{\Gamma(2-2h)}{\Gamma(1-h+im-s)}(im)^{h-1+im-s}\times \label{R^far at infinity s}\\
 &&~~~\times\left[1- \frac{(-im)^{2h-1}}{(im)^{2h-1}}\frac{\sin\pi(h+im)}{\sin\pi(h-im)}\right]r^{-1+im-2s}e^{imr/2} \,,\notag\\
\hat R_{\ell m \hat\omega}^{far}(r\to 0)&=&P\, r^{h-1-s}+Q\,r^{-h-s}\,,\label{R^far at throat s}
\end{eqnarray}
where \be\label{P/Q s}
{P\over Q}= -(-im)^{2h-1}\frac{\Gamma(2-2h)}{\Gamma(2h)}\frac{\Gamma(h-im+s)}{\Gamma(1-h-im+s)}\,.
\ee
To match this with a solution of the near-NHEK equation,
\bea\label{nearNHEK radial eqn s=-2}
&&r(r+2\kappa)\hat R_{\ell m \hat\omega}'' + 2(s+1)(r + \kappa)\hat R_{\ell m \hat\omega}' \\ &&~~+\left[\frac{(mr+n\kappa)^2-2is(r+\kappa)(mr+n\kappa)}{r(r+2\kappa)}+2ism+m^2+s(s+1)-K_\ell\right]\hat R_{\ell m \hat\omega} = \hat T_{\ell m \hat\omega}\,,\notag
\eea
where the source $\hat T_{\ell m \hat\omega}$ is due to the plunging particle on (\ref{t(r) plunge}-\ref{phi(r) plunge}), we need only adjust the Neumann boundary conditions of the solution found in section \ref{section: s=-2 gravity analysis} to the leaky boundary conditions given in \eqref{R^far at throat s}. This is done by adding an ingoing at the horizon homogeneous solution of \eqref{nearNHEK radial eqn s=-2},
\be
\hat R_{in}^{near}=r^{-in/2-s}\left({r\over 2\kappa} +1\right)^{i(n/2-m)-s} {_2}{F}{_1}\left(h-im-s\,,1-h-im-s\,,1-in-s\,,-{r\over 2\kappa}\right)\,,
\ee
with the appropriate amplitude to match the Dirichlet mode of \eqref{R^far at throat s}. Doing so we find (note that $\hat\psi^{(-2)}=M^6 \psi^{(-2)}$):
\bea
&&Q=2M^7\mathcal{N} \times\\
&&~~\times\left[1-(-2im\kappa)^{2h-1}\frac{\Gamma(1-2h)^2}{\Gamma(2h-1)^2} \frac{\Gamma(h-im-s)\Gamma(h-im+s)}{\Gamma(1-h-im-s)\Gamma(1-h-im+s)} \frac{\Gamma(h-i(n-m))}{\Gamma(1-h-i(n-m))}\right]^{-1}\,,\notag
\eea
where $\mathcal{N}$ is given in \eqref{N expression s}. Plugging into equation \eqref{R^far at infinity s} we obtain the full frequency domain solution at asymptotic flat infinity (for $m>0$):
\begin{eqnarray}
&&\hat R_{\ell m \hat\omega}^{far}(r\to \infty) =\frac{m_0 M^2}{6\sqrt{6\pi}}(-1)^{-h}(2\kappa)^{h-1}(3R_0/4)^{i(n-m)}\times \label{R^far at infinity final s}\\
&&~~~~\times \frac{ \gamma \, (im)^{h+in} e^{\pi m} {(2h-1)\Gamma(2+h-im)\Gamma(-2+h-im)}/{\Gamma(2h)^2} }{ \frac{1}{\Gamma(h-i(n-m))} -(-2im\kappa)^{2h-1}\frac{\Gamma(1-2h)^2}{\Gamma(2h-1)^2} \frac{\Gamma(2+h-im)\Gamma(-2+h-im)}{\Gamma(3-h-im)\Gamma(-1-h-im)} \frac{1}{\Gamma(1-h-i(n-m))} }\, r^{3+im}e^{imr/2}\,, \notag
\end{eqnarray}
where
\bea
\gamma&=&2\left[(h^2-h+6-im)S+4(2i+m)S'-4S''\right]W_{im-2,h-\half}(3im/2) \label{gamma definition}\\
&&+\left[(4+3im)S-8iS'\right]W_{im-1,h-\half}(3im/2)\,.\notag
\eea

\subsection{Quasinormal mode decomposition}
As in the scalar case, the QNMs are determined by the zeros of the denominator in \eqref{R^far at infinity final s},
\be
\mathcal{D}=\frac{1}{\Gamma(h-i(n-m))} -(-2im\kappa)^{2h-1}\frac{\Gamma(1-2h)^2}{\Gamma(2h-1)^2} \frac{\Gamma(2+h-im)\Gamma(-2+h-im)}{\Gamma(3-h-im)\Gamma(-1-h-im)} \frac{1}{\Gamma(1-h-i(n-m))}\,.
\ee
This implies that the QNM frequencies are again as in \eqref{qnm frequencies} where $\mathcal{D}$ gives simple poles as in \eqref{D near qnms} with,
\be
\eta_{QNM} =  \frac{\Gamma(1-2h)^2}{\Gamma(2h-1)^2} \frac{\Gamma(2+h-im)\Gamma(-2+h-im)}{\Gamma(3-h-im)\Gamma(-1-h-im)} \frac{(-1)^{N}}{N! \, \Gamma(1-2h-N)}\,. \label{etaQNM s}
\ee
The QNM contribution to the waveform at asymptotic flat infinity is:
\begin{eqnarray}
&&\psi^{(-2)}_{far}(r\to\infty) \notag\\
&&~=\sum_{\ell, m} \frac{3^{h-{3\over 2}}m_0 M}{2^{h+3}} (-1)^{-h} e^{3im/4}\,\gamma\, (im)^{2h+im} e^{\pi m} \frac{(2h-1) \Gamma(2+h-im)\Gamma(-2+h-im)}{\Gamma(2h)^2} \times \notag \\
&&\qquad\qquad\times   \exp \left[ -i\frac{m - i \kappa h}{2M}(\hat{t}-\hat{t}_0)-\frac{3im}{4 \kappa} e^{-\frac{\kappa}{2M} (\hat{t}-\hat{t}_0)} \right] e^{i m (\hat{\phi} - \hat{\phi}_0)}  S_{\ell}(\theta) \, r^{3+im} e^{imr/2} \,,
\end{eqnarray}
where $\gamma$ is given in \eqref{gamma definition}.

\section*{Acknowledgements}

APP and AS are grateful to J. Hewlett for collaboration at an earlier stage. SH acknowledges B. Kol for helpful discussions. APP and AS were supported in part by DOE grant DE-FG02-91ER40654. SH was supported by the Israel Science Foundation grant no. 812/11 and by the Einstein Research Project ``Gravitation and High Energy Physics,'' which is funded by the Einstein Foundation Berlin.

\appendix

\section{Throat geometries in near extreme Kerr}\label{app on throat geometries}

There exist three different scaling limits one can take to zoom into the geometry of the throat of a near extreme Kerr black hole. Consider the Kerr metric \eqref{Kerr metric}. Define
\be
a=M\sqrt{1-(\kappa\delta)^2}\,,
\ee
and make the coordinate transformation:
\be
r=\frac{\hat r-r_+}{\delta^p r_+},\qquad t=\frac{\delta^p \hat t}{2M},\qquad\phi=\hat\phi-\frac{\hat t}{2M}\,,
\ee
where $p>0$. Taking $\delta\to 0$ we have the following three cases.
\\[0.25in]
For $p<1$ we get NHEK:
\be
ds^2=2 M^2 \Gamma (\theta)\left(-r^2 dt^2+\frac{dr^2}{r^2}+d\theta^2+\Lambda (\theta)^2(d\phi+rdt)^2\right)\,.
\ee
For $p=1$ we get near-NHEK:
\be
ds^2=2 M^2 \Gamma (\theta)\left(-r(r+2\kappa) dt^2+\frac{dr^2}{r(r+2\kappa)}+d\theta^2+\Lambda(\theta)^2\left(d\phi+(r+\kappa)dt\right)^2\right)\,.
\ee
For $p>1$ we get horizon-NHEK:
\be
ds^2=2 M^2 \Gamma (\theta)\left(\frac{dr^2}{2\kappa r}+d\theta^2+\Lambda (\theta)^2(d\phi+\kappa dt)^2\right)\,.
\ee
Note that horizon-NHEK is a singular metric because it has vanishing determinant.

\section{Equatorial geodesic equations in NHEK-like metric}\label{app on geodesics}

Consider a NHEK-like metric:
\be\label{metric}
ds^2=2M^2\Gamma(\theta)\left[-N^2dt^2+\frac{dr^2}{N^2}+d\theta^2+\Lambda(\theta)^2(d\phi+N^\phi dt)^2\right]\,,
\ee
where $N^2, N^\phi$ are functions of $r$ only. Confining ourselves to the equatorial plane, $\theta=\pi/2$, the two constants of motion associated with the Killing fields $\p_t$ and $\p_\phi$, namely the energy and angular momentum (per unit rest mass) $E$ and $L$, suffice to describe the geodesics. The timelike equatorial geodesics, parameterized by proper time $\tau$, are given by:
\begin{eqnarray}
\frac{dt}{d\tau}&=&\frac{E+LN^\phi}{M^2N^2}\,,\label{eq t}\\
\frac{d\phi}{d\tau}&=&\frac{L}{4M^2}-N^\phi\cdot\frac{E+LN^\phi}{M^2N^2}\,,\label{eq phi}
\end{eqnarray}
and for the radial motion:
\be
\Sigma^2\left(\frac{dr}{d\tau}\right)^2=V(r)\,,\label{eq r}
\ee
with $\Sigma^2=M^2/N^2$ and,
\be
V(r)=\frac{4E^2+8ELN^\phi-4M^2N^2+4L^2(N^\phi)^2-L^2N^2}{4M^2N^2}\,.
\ee
For a circular orbit at some radius $r$ we need $V(r)=0$ and $V'(r)=0$. These equations can be solved to give $E$ and $L$ in terms of the circle radius but circular orbits don't necessarily exist for all values of $r$. Moreover, not all circular orbits are stable. Stability requires $V''(r)\leq 0$.

\subsection{near-NHEK}
For near-NHEK:
\be
N^2=r(r+2\kappa)\,,\quad N^\phi=r+\kappa\,.
\ee
Circular orbits, satisfying $V(r)=0$ and $V'(r)=0$, exist for $3r^2+6\kappa r-\kappa^2>0$ and have:
\begin{eqnarray}
E&=&\mp\frac{2M\kappa^2}{\sqrt{3r^2+6\kappa r-\kappa^2}}\,,\\
L&=&\pm\frac{2M(r+\kappa)}{\sqrt{3r^2+6\kappa r-\kappa^2}}\,,
\end{eqnarray}
where the upper (lower) sign refers to direct (retrograde) orbits. Among those, stable ones should also have $V''(r)\leq 0$ which is equivalent to:
\be
\frac{8\kappa^2}{r(r+2\kappa)(3r^2+6\kappa r-\kappa^2)}\leq 0\,.
\ee
Thus in near-NHEK ISCO is at $r=\infty$. Since ISCO is a marginally stable orbit, we choose the plunging orbit with $E$ and $L$ those of ISCO (direct orbit):
\be\label{E and L for plunge orbit}
E=0\,,\qquad L={2M \over \sqrt{3}}\,.
\ee
Plugging these values into equations (\ref{eq t}--\ref{eq r}) we can integrate them to find the expressions for the plunge trajectory in near-NHEK:
\begin{eqnarray}
t(r)&=&\frac{1}{2\kappa}\ln\frac{1}{r(r+2\kappa)}+t_0\,,\\
\phi(r)&=&\frac{3r}{4\kappa}+\frac{1}{2}\ln\frac{r}{r+2\kappa}+\phi_0\,.
\end{eqnarray}

\subsection{NHEK}
For NHEK:
\be
N^2=r^2\,,\quad N^\phi=r\,.
\ee
Circular orbits, satisfying $V(r)=0$ and $V'(r)=0$, exist at all $r>0$ and they all have:
\begin{eqnarray}
E&=&0\,,\\
L&=&\pm\, {2M \over \sqrt{3}}\,.
\end{eqnarray}
Moreover, all of them are marginally stable with $V''(r)=0$. So there is no single \emph{innermost} stable circular orbit (ISCO) in NHEK. Since these are marginally stable orbits the solution to the equatorial geodesic equations with $E=0, L=2M/\sqrt{3}$ may be called a (direct) plunge trajectory in NHEK. Plugging these values into equations (\ref{eq t}--\ref{eq r}) we can integrate them to find the expressions for the plunge trajectory in NHEK:
\begin{eqnarray}
r(t)&=&r_0\,,\\
\phi(t)&=&-\frac{3}{4}r_0\,t+\phi_0\,.
\end{eqnarray}

\end{document}